\documentclass[12pt]{article}
\usepackage{a4,graphicx,epsfig}
\def\question#1 {~\\ {\bf\it #1 }\\}

\newcommand{\be}{\begin{equation}}
\newcommand{\beq}{\begin{equation}}
\newcommand{\ee}{\end{equation}}
\newcommand{\eeq}{\end{equation}}
\newcommand{\ba}{\begin{eqnarray}}
\newcommand{\ea}{\end{eqnarray}}
\newcommand{\bb}{\begin{thebibliography}}
\newcommand{\eb}{\end{thebibliography}}

\newcommand{\lab}[1]{\label{#1}}

\begin{document}

\title{ Unitarisation of the hard pomeron \\ and black-disk limit
at the LHC} 
\vskip 1cm
\author{J.-R. Cudell\footnote{Institut de Physique, B\^at. B5a, Universit\'e de Li\`ege, Sart
Tilman, B4000
  Li\`ege, Belgium, e-mail: J.R.Cudell@ulg.ac.be },
and O.V. Selyugin\footnote{Bogoliubov
 Laboratory of Theoretical Physics, JINR, 141980, Dubna, Moscow Region,
 Russia, e-mail: selugin@theor.jinr.ru.}}
\maketitle
\vskip 1cm
\begin{quote}
\centerline{\bf Abstract}
{\small\noindent
Recent models of soft diffraction include a hard pomeron pole besides
the usual soft term. Such models violate the black-disk limit
around Tevatron energies, so that they need to be supplemented by
a unitarisation scheme. Several such schemes are considered
in this letter, where we show that they lead to a large uncertainty 
at the LHC. We also examine the impact of unitarisation on various
small-$t$ observables, the slope in $t$
of the elastic cross section, or the ratio of
  the real to imaginary parts of the scattering amplitude, leading to
the conclusion that the existence of a hard pomeron in soft scattering
may be confirmed by LHC data.}
\end{quote}

\vskip 2cm

\section*{Introduction: hard poles}
Experimental data reveal that total cross sections
grow with energy. This means that the leading contribution in the
high-energy limit is given by the rightmost singularities in the complex-$j$
plane, the pomerons,
with intercepts exceeding unity. In the framework of perturbative QCD,
the leading singularity is expected to exceed unity by an amount
proportional to $\alpha_s$ \cite{lipatov1}. At leading-log, one obtains
a leading singularity at $J-1 = 12 \alpha_s\ \log 2/\pi$.
From analytic $S$-matrix theory, it is expected that the singularities
of the exchanges do not depend on the external kinematics. Hence, although
perturbative calculations can be justified only for the scattering of 
far off-shell particles, the singularities that they predict
 should remain valid in the long-distance
range.

In a recent study \cite{clms}, we have indeed found that forward data 
(total cross sections and the ratios of the real part to the imaginary part of
the amplitude) could be fitted well by a combination of a soft pomeron
(which would be purely non perturbative)
and a hard pomeron. 
The expression of the leading terms of the
total cross sections  for the scattering of $a$ on $p$ becomes
 \begin{equation}
\label{hardpompole}
\sigma _{tot}^{ap}=\frac{1}{2P\sqrt{s}}\Im mA
\left({s-u\over 2}\right)
\end{equation}
with $P$ the beam momentum in the target frame, $s$, $t$ and $u$
the Mandelstam variables, and $A$ the hadronic amplitude 
\begin{equation}
\label{poles}
\Im mA(s)\approx {H_a}\left( \frac{s}{s_{1}}\right) ^{\alpha _{H}(0)}
+{S_a}\left( \frac{s}{s_{1}}\right) ^{\alpha _{S}(0)}
\end{equation}
with \( s_{1}=1 \) GeV\( ^{2} \), and the parameters given in Table~1.
\begin{table}
{\begin{center} \begin{tabular}{||c|c|c||c|c|c||}
\hline\hline
parameter & value & error&parameter & value & error \\ \hline\hline
\( \alpha _{S}(0) \) & 1.0728 & 0.0008 &\( \alpha _{H}(0) \) & 1.45 & 0.01  \\ \hline
\( S_{p} \)& 56.2 & 0.3  &\( H_{p} \)& 0.10 & 0.02\\ \hline
\( S_{\pi } \) & 32.7 & 0.2  &\( H_{\pi } \)& 0.28 & 0.03\\ \hline
\( S_{K} \)& 28.3 & 0.2  &\( H_{K} \)& 0.30 & 0.03 \\ \hline
\( S_{\gamma } \)& 0.174 & 0.002  &\( H_{\gamma } \)& 0.0006 & 0.0002\\ \hline\hline
\end{tabular}\end{center}}
\label{parameters}
\caption{Parameters of the leading singularities of the fits of ref.~\cite{clms}
for \protect\( \sqrt{s}\protect \) from 5 to 100~GeV. }
\end{table}
The inclusion of these two pomerons, together with the use of
integral dispersion relations, and the addition of sub-leading
meson trajectories ($\rho$/$\omega$ and $a/f$), leads to
a successful description of all $pp$, $\bar p p$, $\pi^\pm p$,
$K^\pm p$, $\gamma p$ and $\gamma\gamma$ data for $\sqrt{s}\leq 100$ GeV.

The problem however is that such a fit does not extend to 
high energies. Indeed, the fast growing hard pomeron leads to a violation
of unitarity (an elastic cross section bigger than the total cross section)
for values of $s$ smaller than 1 TeV.

In \cite{clms}, we used a simple ansatz to unitarise the hard pole. However,
we did not examine the uncertainties linked to it, as we simply wanted to
show that it was possible to extend the fit to higher energies.
In this letter, we propose to reconsider this question
and to give an estimate of the uncertainties on total cross section
measurements at the LHC.

Section 1 will be devoted to a reminder of unitarisation,
and of the impact-parameter ($\vec b$) formalism. 
Section 2 will consider the ``minimal"
unitarisation, which cuts off the amplitude in impact parameter once it
reaches the black-disk limit.
Section 3 will consider analytic 
unitarisation schemes. 
Finally, putting everything together,
we shall show that the cross section at the LHC should be large, of the
order of 150 mb.
\section{Unitarity and black-disk limit}
The simplest expression of the unitarity constraints can be obtained from
partial-wave amplitudes. We can write
\beq A(s,t)=8\pi\sum_l (2l+1){\cal F}_l(s) P_l(\cos\theta_s).\eeq
At high energy and small angle, one can rewrite $l\approx {b\sqrt{s}\over 2}$,
so that the partial-wave decomposition can be rewritten in impact-parameter space as
\newcommand{\F}{{\cal F}}
\beq
A(s,t=-q^2)=2s\int d^2\vec b e^{-i \vec q.\vec b} \F(s,b)
\eeq
In terms of $\F(s,b)=\int d^2\vec q e^{i \vec q.\vec b} A(s,t)/(8\pi^2 s)$, the total and elastic cross sections are given by
\beq
\sigma_{tot}=4\pi\int bdb\Im m\F(s,b)
\label{tot}
\eeq 
and 
\beq 
\sigma_{el}=2\pi\int bdb|\F(s,b)|^{2}.
\label{el}
\eeq

One can then show that 
unitarity of the $S$ matrix, $S S^\dagger=1$, 
together with analyticity, and with the normalisation used in (\ref{tot})
and (\ref{el}), requires that
\beq
0\leq\left|\F(s,b)\right|^{2}\leq 2\Im m \F(s,b)\leq 4.
\label{unitarity}
\eeq
   At high energies, as $\Im m \F\propto s^\Delta$ with $\Delta>0$, 
the scattering amplitude $\F(s,b)$ 
reaches the unitarity bound for some value $b_u(s)$ of the
impact parameter. 

However, before this happens, another regime is reached:
one gets the maximum inelasticity if $2\Im m \F(s,b)-|\F(s,b)|^2$ is
maximum, i.e. if 
\beq
\Im m \F(s,b)=1.\eeq
 This is the region where the proton
becomes black, and it is usually referred to as the black-disk limit. The
imaginary part of $\F$ is usually noted as $\Gamma(s,b)$ and called the
profile function. 
Saturation usually refers to the black-disk limit. 
We shall use it to mean that we reach the maximum possible 
inelastic amplitude at some distance $b_S(s)$
between scattering particles. 

In QCD, this can be thought of
as a consequence of the growth of the gluon density at small $x$, which
must be tamed by non-linear effects
connected to the next-to-leading terms.
One needs to note that the role of  non-perturbative effects 
connected with confinement \cite{kovner-cf} and with large impact
parameters makes the analogy with perturbative saturation questionable,
as in the present work saturation will first occur at small $b$.

One expects  saturation
 to tame the growth of  $\sigma_{tot}$.
   For example,
in  the loop-loop correlation model (LLCM) \cite{sochi}
 based on the functional integral approach to high-energy
collisions \cite{nach},
the $T$-matrix element for elastic
proton-proton scattering
reads
\begin{eqnarray}
        A_{pp}(s,t)
        & \propto & \,\,
        i s \ \int  \ d^2\vec b\,
        e^{i {\vec q}.{\vec b}}\,
        \Gamma(s,b)
\label{Eq_T_pp_matrix_element} \nonumber \\
        \Gamma(s,b)
        & = &
        \int \!\!dz_1 d^2r_1\!\! \int \!\!dz_2 d^2r_2
        |\psi_p(z_1,\vec{r}_1)|^2 |\psi_p(z_2,\vec{r}_2)|^2
\nonumber\\ &&
        \times
        \left[1-S_{DD}(s,{\vec b},z_1,{\vec r}_1,z_2,{\vec r}_2)
      \right]
\label{Eq_model_pp_profile_function}
\end{eqnarray}
with 
$\psi$ the proton light-cone wave function
and $S_{DD}$ the dipole-dipole S matrix element.
The saturation of the profile function is a direct consequence
of $S$-matrix unitarity.
 In this model, the saturation
  regime
  at small $b$ is reached only at very small $x \approx 10^{-10}$ and very high
  energies $\sqrt{s} \geq 10^{6} \ $GeV, where 
there is a transition from a power-like to an
$\log^2$-increase of $\sigma^{tot}_{pp}(s)$, which then respects the
Froissart bound \cite{Froissart}.
A similar result, in the framework of the dipole picture of soft processes,
also leads to a taming $\sigma_{tot}$
but at  lower energy \cite{bart}.

If we take a single simple pole for the scattering amplitude,
with an exponential form factor of slope $d$,
\beq
A(s,t)=S_p s^{\alpha_S(0)+\alpha' t} e^{dt}
\eeq
then the radius of saturation and its dependence on energy
 can be obtained analytically:
\begin{equation}
  b_S(s)^2 =4(d+\alpha' \log s )
\log\left({S_p s^{\alpha_S(0)-1}\over 2(d+\alpha' \log s )}\right)
\end{equation}
Approximating $\sigma_{tot}\approx 2\pi b_S^2(s)$, one sees that
the total cross section grows logarithmically at medium energies
and like $\log^2 s$ at very high energies.

In our model \cite{clms,clm}, the $pp$ elastic scattering amplitude is
  proportional to the hadrons form-factors and can be approximated
  at small $t$ as:
\begin{eqnarray}
 A(s,t)  &=&  [ H_p\ F_H(t) (s/s_1)^{\alpha_H(0)}
           e^{\alpha^{\prime}_H \  t \ \log (s/s_1)}\nonumber\\
          &+& S_p\ F_S(t) \ (s/s_1)^{\alpha_S(0)}
             e^{\alpha^{\prime}_S \  t \ \log (s/s_1)} ]
\end{eqnarray}
where the couplings and intercepts are given in Table 1.
The study of ref.~\cite{clm} of small-$t$ elastic scattering shows 
that the slope $\alpha'_S$ of the soft pomeron trajectory
is slightly higher than its 
classical value \cite{DL,book}, and we shall take $\alpha'_S=0.3$ GeV$^2$.
The slope of the hard pomeron trajectory is evaluated \cite{clm,DLH} to be
$\alpha'_H=0.1$~GeV$^2$.
The normalisation $s_1=1$ GeV$^2$ will be dropped below and
$s$ also contains implicitly the phase factor $\exp(-i \pi/2)$, corresponding to
crossing symmetry.

A small-$t$ analysis \cite{clm} indicates that the form factor
$F_S(t)$ is  close to the square of
 the Dirac elastic form factor,
and can be approximated by the sum of three exponentials \cite{book}.
 \begin{eqnarray}
  F_S(t)&=&\left(\frac{4 m_p^2-2.79 t}{4 m_p^2-t} 
\frac{1}{1-t/\Lambda^2}\right)^2\nonumber\\
 &\approx& h_{1} e^{d_1 \ t} \ + \  h_{2} e^{d_2 \ t} \
 + h_{3} e^{d_3 \ t}.   
\label{formfac}
 \end{eqnarray}
where $m_p$ is the mass of the proton, $\Lambda^2=0.71$ GeV$^2$.
The other parameters are given in Table~2.
\begin{table}
{\begin{center} \begin{tabular}{||c|c||c|c||}
\hline\hline
parameter & value &parameter & value (GeV$^{-2}$) \\ \hline\hline
$h_{1}$   & 0.55  & $d_1$ & 5.5 \\\hline
$h_{2}$   & 0.25  & $d_2$ & 4.1 \\\hline
$h_{3}$   & 0.20  & $d_3$ & 1.2\\\hline\hline
\end{tabular}\end{center}}
\label{tform}
\caption{Parameters of the elastic pomeron form factor, see Eq.~(\ref{formfac}).}
\end{table}
For the hard pomeron, the form factor is rather uncertain \cite{clm},
and we assume that it can be taken equal to $F_S(t)$.

\begin{figure}
\begin{center}
\mbox{\epsfxsize=100mm\epsffile{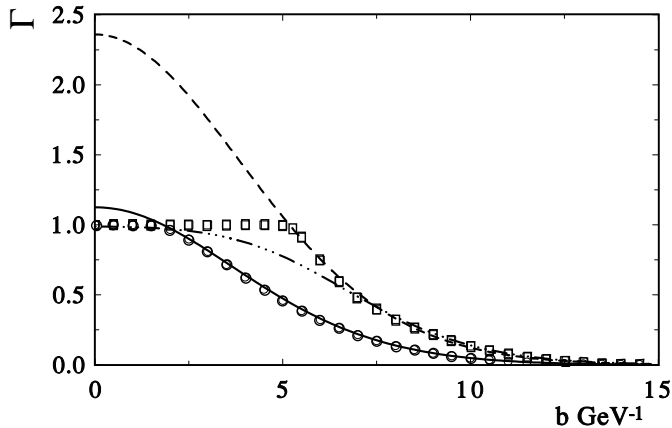}}
\end{center}
\caption{
  The profile function for proton-proton scattering:
the hard line and circles  -  at  $\sqrt{s}=2$~TeV
       without and with the saturation; the  dashed line and squares -
    at  $\sqrt{s}=14$~TeV  without and with saturation.
    the dash-dotted line - the eikonal form (\ref{eik}) at
   $\sqrt{s}= 14$~TeV.
  }
\label{profile}
\end{figure}

We then obtain in the impact parameter representation
a specific form for the amplitude in $\vec b$ space, $\F_0(s,b)$ \cite{dif04},
which we show in Fig.~\ref{profile}:
\begin{eqnarray}
 \F_0(s,b) &= &{S_p\over s}\sum_i {2h_i\over r_{i,S}} 
s^{\alpha_S(0)}
\exp(-b^2 /r_{i,S}^2) + (S\rightarrow H)\nonumber\\
{\mathrm{with}\ }    r_{i,S}^2 &=& 4 \ (d_i + \alpha^{\prime}_S \ \log (s)) \\ 
    r_{i,H}^2 &=& 4 \ (d_i + \alpha^{\prime}_H \ \log (s)).
\label{eqprofile}
\end{eqnarray}
One can see that at some energy and at small $b$,  $\Gamma_0(s,b)=
\Im m\F_0(s,b)$
 reaches  the black disk limit.
For our model,   this will be in the region
$\sqrt{s} \approx 1.5 \ $TeV.

What happens then is largely unclear. There is  no unique procedure
to saturate the amplitude, and saturation and unitarisation will 
be important issues at the LHC. In the following, we shall
consider two main schemes: the minimal one, where the saturation
freezes the profile function at 1: this is the minimum possibility
for the restoration of unitarity, as it does not affect in any way the
low-energy data. In the next section, we shall also
consider the predictions of an eikonal scheme.

Saturation of the profile function will surely control
the behaviour of $\sigma_{tot}$ at higher
energies. We assume that once it reaches 1, the amplitude
does not change anymore and remains equal to $i$: 
recombination must be maximal for black
protons. But this freezing of the profile function must be implemented
carefully: one cannot simply cut the profile function sharply
as this would lead to a non-analytic amplitude, and to specific 
diffractive patterns in the total cross section and in the slope of
the differential cross sections. Furthermore, we have to match at
 large impact parameter the behaviour of the unsaturated
profile function.

We use an analytic interpolating function, which is equal to one
for large impact parameters
and which forces the profile function to approach 1 at
the saturation scale $b_s$ as a Gaussian. Analyticity of the
function enables us to use a complex $s$ as before to obtain the real
part.

We assume saturation starts at a point $b_0$ a little before $b_s$, and that
the profile function is 1 for $b<b_0$. 
The saturated profile, $\Gamma_s(s,b)$,
is otherwise given by
\beq \Gamma_s(s,b)={\Gamma\left(s, b-{b_0\over 1+ ((b-b_0)/b_z)^2}\right)\over
1  +  (\Gamma(s,0)-1) \exp\left[-\left({\left({(b-b_0)\over b_y(s)}\right)}\right)^2
\right]   }
\eeq
We find that we need to assume that the scale in the Gaussian is $s$-dependent
because the slope of the profile function decreases with energy. A
reasonable match is provided by $b_y=32/ \log s$, which is about 5 GeV$^{-1}$
at the Tevatron. We shall give our results for $b_z^2=2$ GeV$^{-2}$ and
$b_0/b_s=97.5 \%$. 

\begin{figure}
\begin{center}
\mbox{\epsfxsize=100mm\epsffile{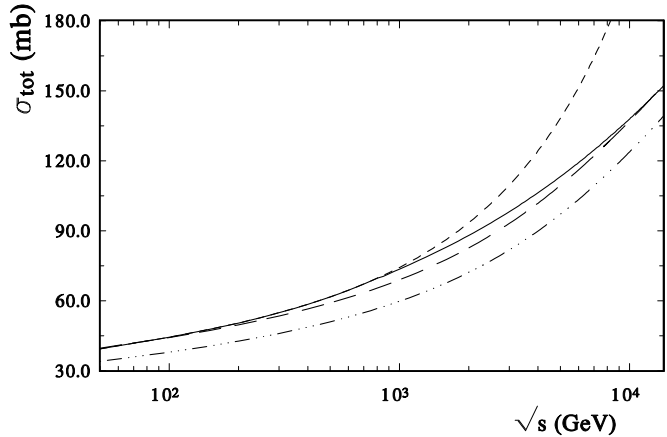}}
\end{center}
\caption{The total cross section as a function of $\sqrt{s}$, 
for the bare amplitude (short dashes), the saturated amplitude (plain curve),
the eikonalised amplitude (dash-dot-dot), and for a renormalised eikonal 
(long dashes).}
\label{sigtot}
\end{figure}

\begin{figure}
\begin{center}
\mbox{\epsfxsize=70mm\epsffile{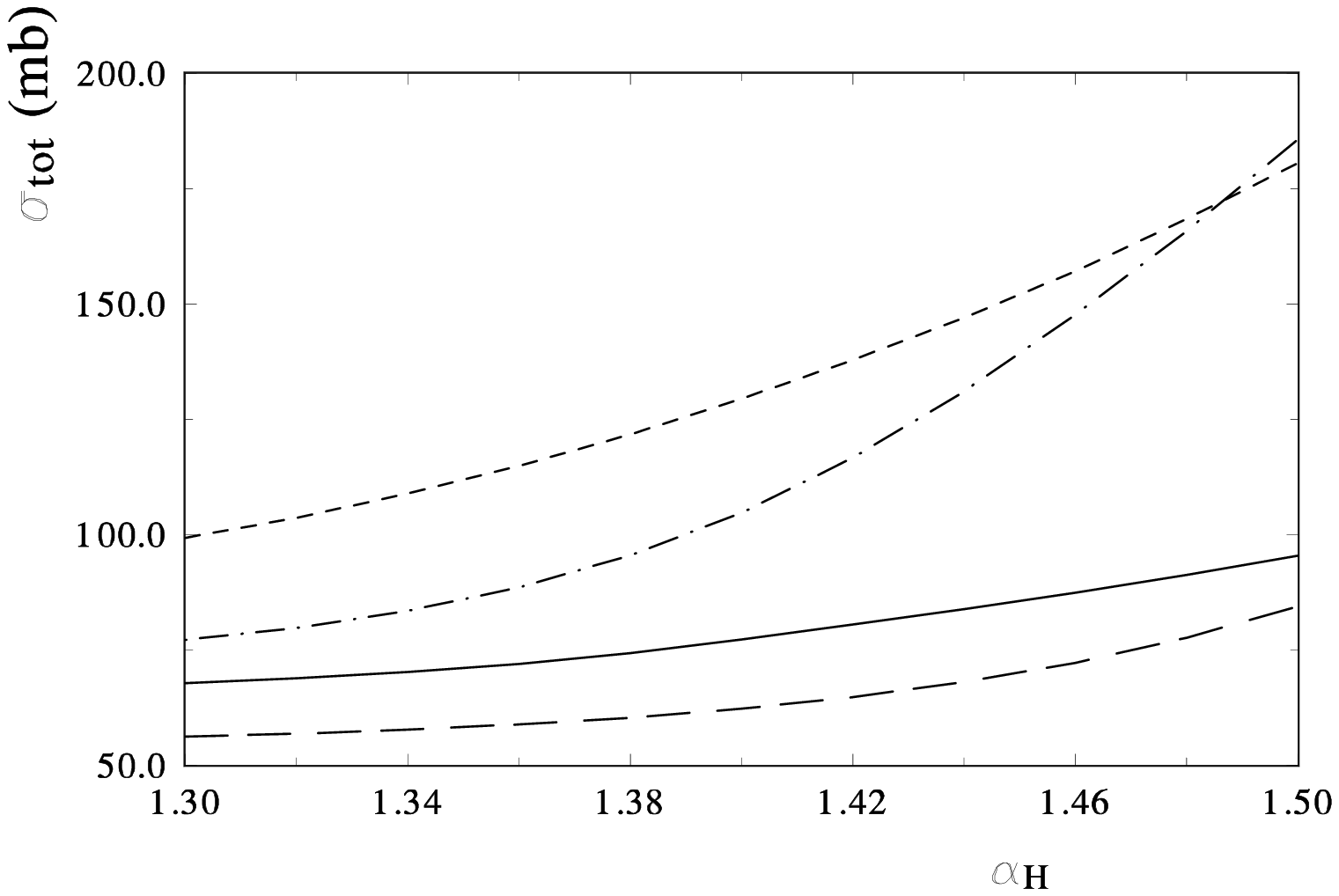}}
\mbox{\epsfxsize=70mm\epsffile{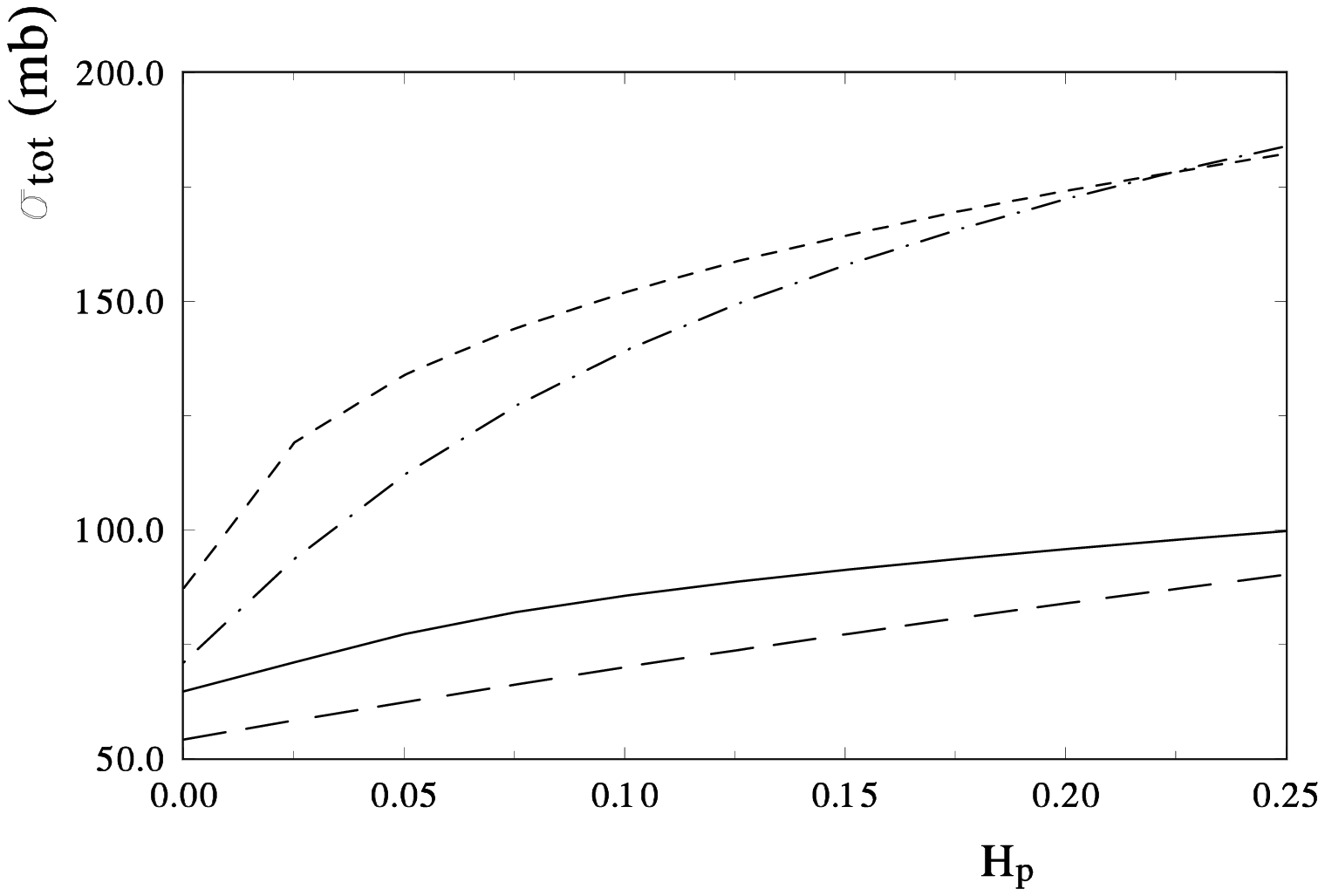}}
\end{center}
\caption{The total cross section at the Tevatron (lower curves) and at the LHC
(upper curves) as a function of the intercept (left) and of the coupling (right)
of the hard pomeron for a saturated amplitude (plain curves and short dashes)
and for an eikonalised amplitude (long dashes and dash-dots)
.}
\label{coupeps}
\end{figure}

We show in Fig. \ref{sigtot} the  behaviour of the total cross section
at high energies. We see that saturation brings in a significant decrease of the LHC cross section. However, it is also clear that the simple saturation
considered here is not enough, as the total cross section at the Tevatron will
be 85 mb, which is 2 standard deviations from the CDF result. It may be
that our previous estimate of the pomeron coupling or intercept is too big. 
We show in Fig.~\ref{coupeps} the effect of changing either. We see that if
saturation is the driving mechanism, then the coupling would need to be 
reduced\footnote{Note that this is still four times larger than the 
value advocated in \cite{peter}.} 
to 0.06  or the intercept to 1.41 in order to accommodate the Tevatron point in
this scheme. As we want to show qualitatively what the effect of saturation
might be, we keep the parameters of Table~1 in the following.

\begin{figure}
\begin{center}
\mbox{\epsfxsize=70mm\epsffile{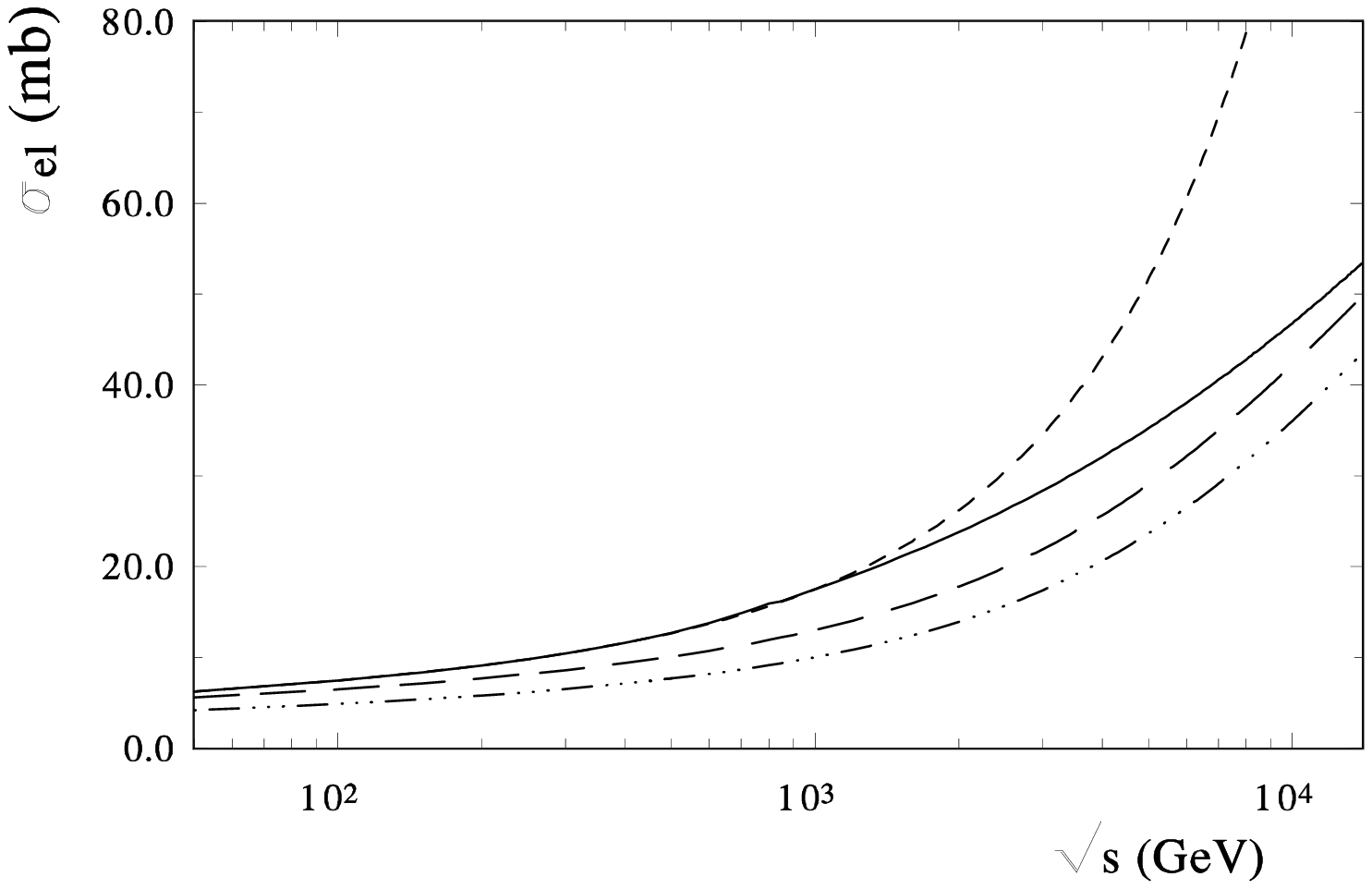}}
\mbox{\epsfxsize=70mm\epsffile{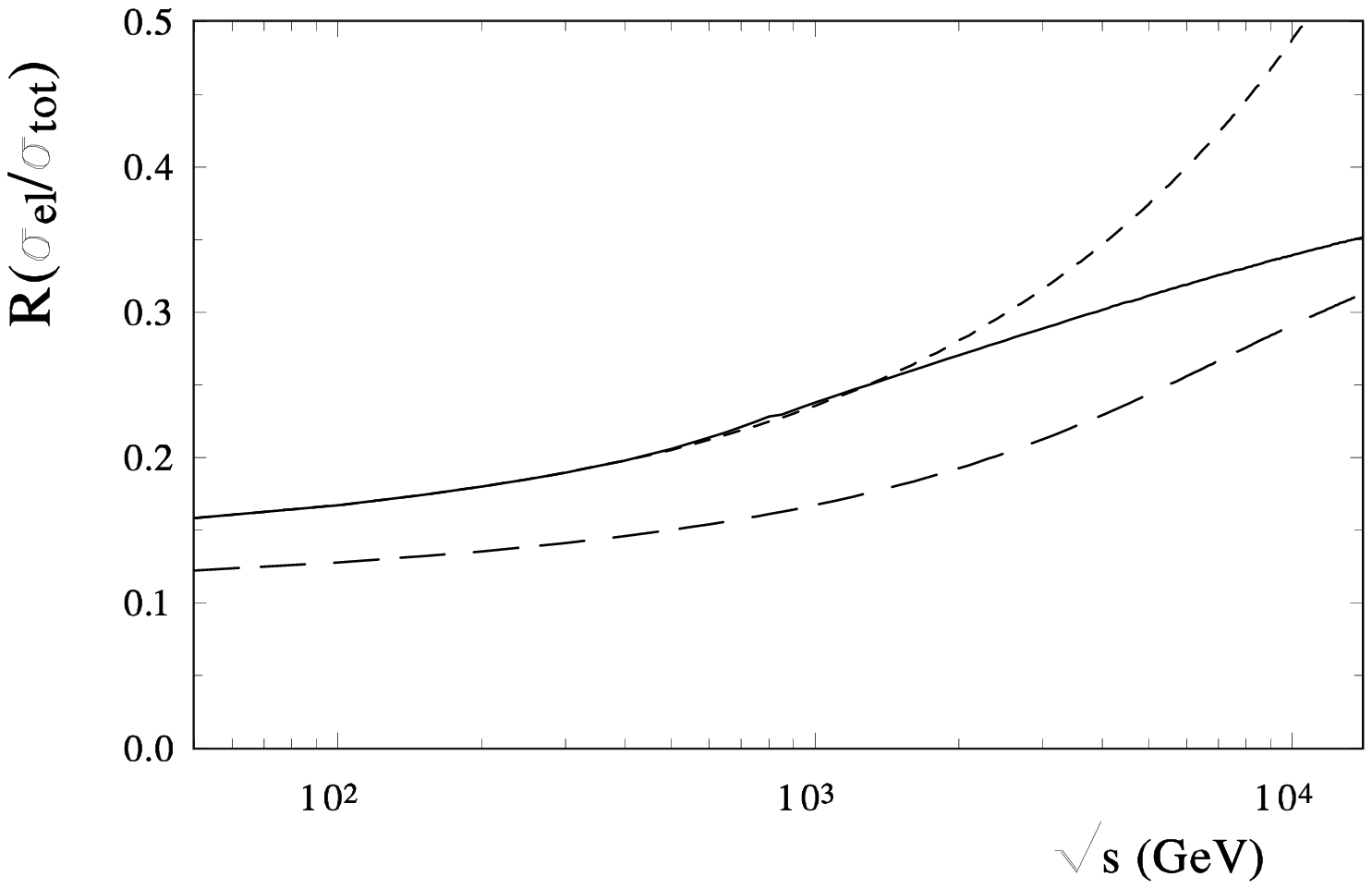}}
\end{center}
\caption{The elastic cross section 
as a function of $s$ (left), and its ratio to the total cross section (right), 
for the bare amplitude (short dashes), the saturated amplitude (plain curve),
the eikonalised amplitude (dash-dot-dot), and for an extended eikonal 
(long dashes).
  }
\label{elastic}
\end{figure}
\begin{figure}
\begin{center}
\mbox{\epsfxsize=70mm\epsffile{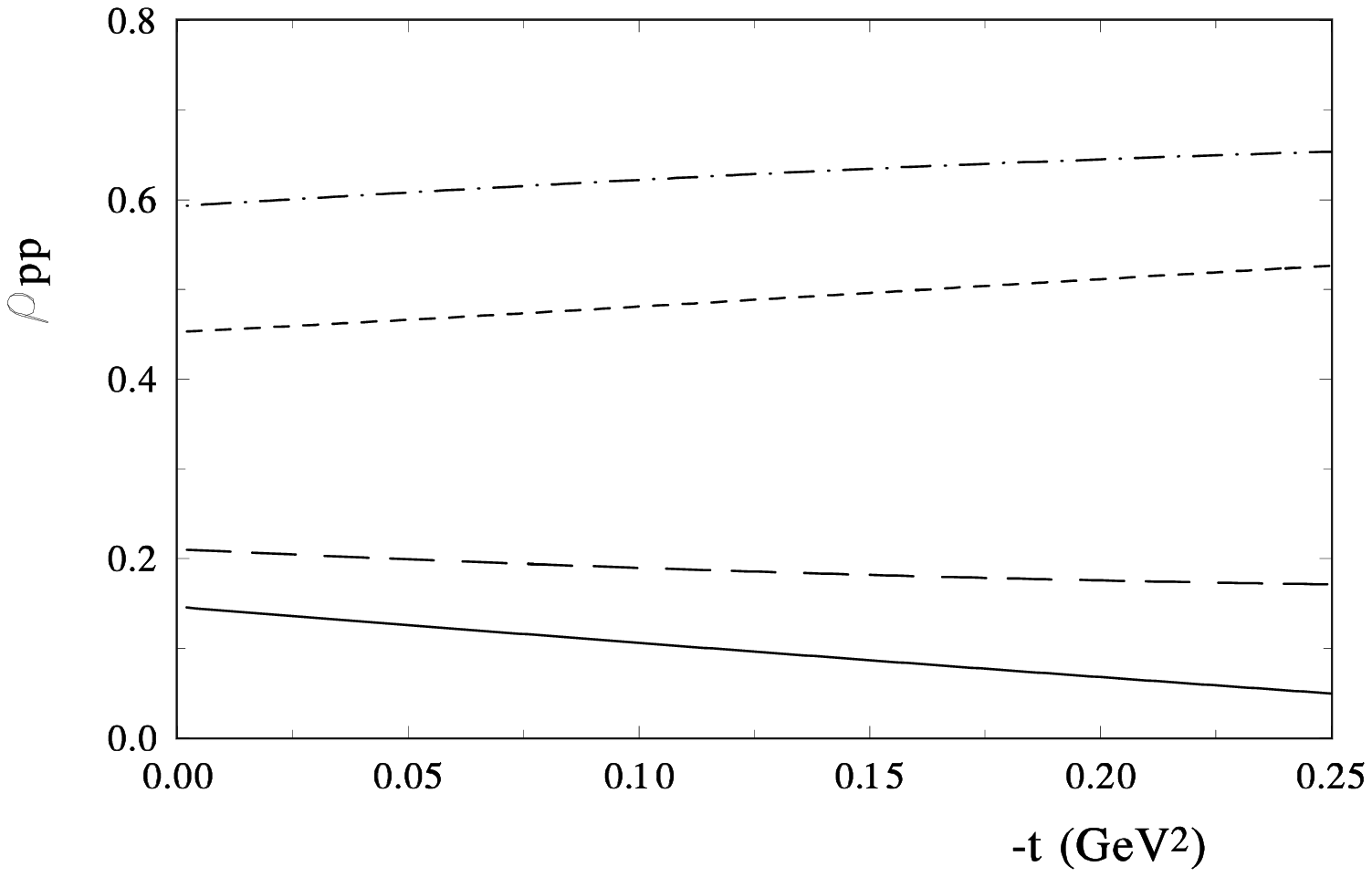}}
\mbox{\epsfxsize=70mm\epsffile{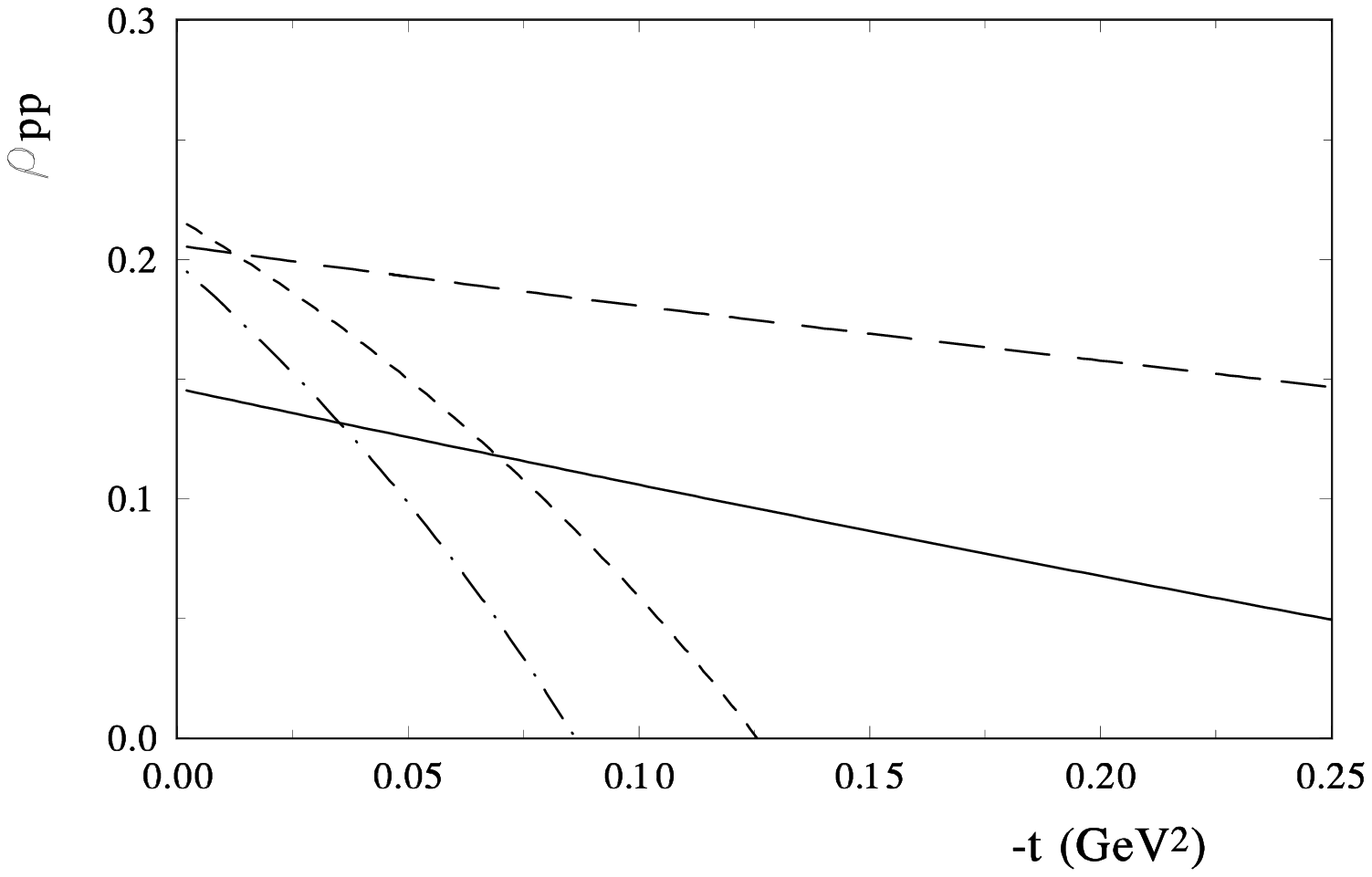}}
\end{center}
\caption{The ratio of the real to the imaginary part of the amplitude as a function of $t$, 
for the bare and the saturated amplitudes 
at various energies: 100~GeV (plain curve),
500~GeV (long dashes), 5~TeV (short dashes) and 14~TeV (dash-dotted curve).
  }
\label{rho}
\end{figure}
\begin{figure}
\begin{center}
\mbox{\epsfxsize=70mm\epsffile{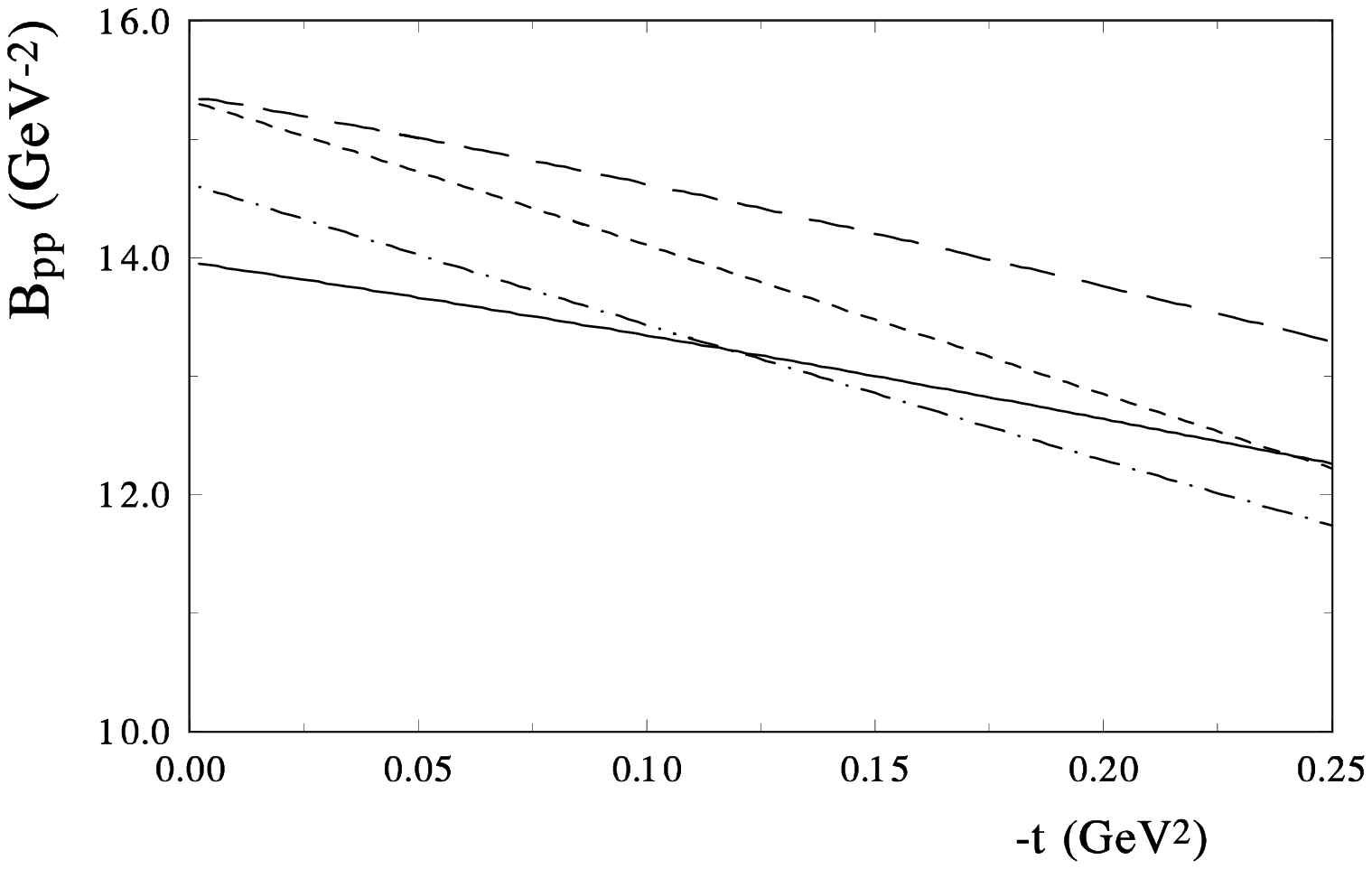}}
\mbox{\epsfxsize=70mm\epsffile{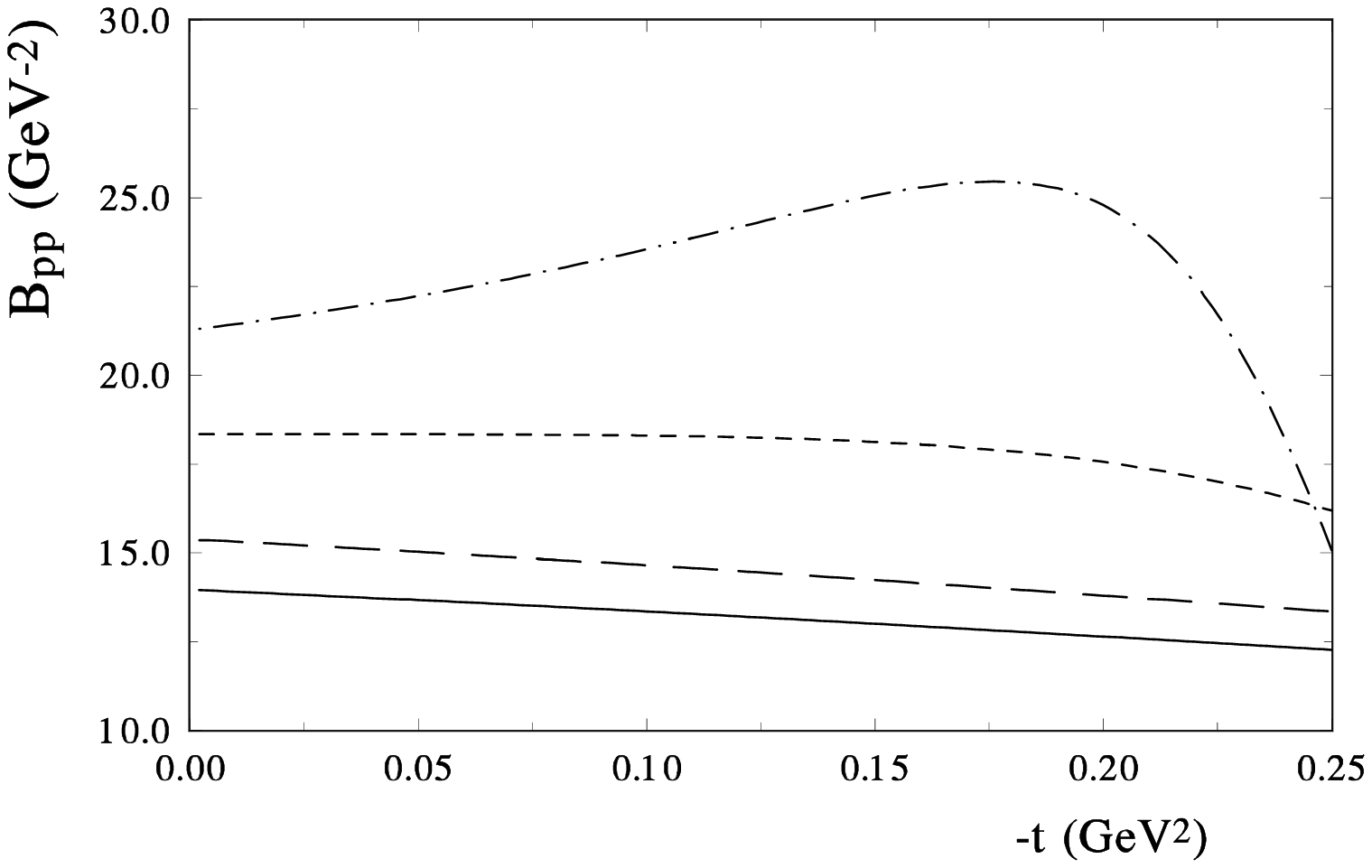}}
\end{center}
\caption{The slope of the elastic differential cross section 
as a function of $t$, 
for the bare and saturated amplitudes at various energies: 100 GeV (plain curve),
500 GeV (long dashes), 5 TeV (short dashes) and 14 TeV (dash-dotted curve).
  }
\label{slopeel}
\end{figure}

The saturation regime will have some major effects at the LHC. We show in Fig.~\ref{elastic} that the elastic cross section will be somewhat affected, and
that its growth will be tamed: the ratio $\sigma_{el}/\sigma_{tot}$ will
start a slow growth towards 0.5. But more importantly, the small-$t$ data
will look quite different. We show in Fig.~\ref{rho} the behaviour of the
ratio of the real-to-imaginary part of the cross section, both in the bare
and in the saturated case. From it, we see that the small-$t$ slope of $\rho$
will be one of the most striking features of saturation.
If this is not measurable at the LHC, then one can also consider the slope of
the differential elastic cross section, which we show in Fig.~\ref{slopeel}.
We see that saturation increases the slope at small $t$, and predicts a
fast drop around $|t|=0.25$ GeV$^2$, when one enters the region of the dip.

In fact, saturation naturally predicts 
a small increase of the slope with $t$ at small
$t$. To understand this,
let us take the simple form of the black disk with a sharp edge at radius $R$.
  The scattering
  amplitude can then be represented as
\begin{eqnarray}
    A(s, t\!=\!0) \ \sim \ \frac{J_1(\sqrt{-t} R)}{\sqrt{-t} R}\ ,
    \lab{hbd}  \nonumber
\end{eqnarray}
  In this case, the slope of the differential cross section  at small
  momentum transfer will be
\begin{eqnarray}
  B_{BDL} \sim R^2/4 \ + R^4/32 \ |t|.  \lab{b-bdl}
\end{eqnarray}
 Hence the slope will  grow with increasing $|t|$
 at small momentum transfer, as can be seen in Fig.~\ref{slopeel}.

\section{Analytic unitarisation schemes}
As we have seen, in $\vec b$ space, unitarisation can be written
\beq
2\Im m\F-(\Im m\F)^2-(\Re e\F)^2=g_{in}>0.
\eeq
The general solution can be obtained iff $g_{in}<1$, 
and can be written 
$\F=i(1-(1-g_{in})e^{i\Phi}).$
We can rewrite it using the opacity $\Omega$ so that $(1-g_{in})=e^{-\Omega}$
\beq 
\F=i\left(1-e^{-\Omega+i\Phi}\right)=i\left(1-e^{i\chi(s,b)}\right).
\label{eiko}
\eeq
Any unitarisation method has to lead to such a form for the amplitude.
The ambiguity however comes when one tries to identify $g_{in}$ and
$\Phi$ in formula (\ref{eiko}) with the physics input. It is known
from potential models that in non-relativistic physics 
one can think of the Taylor expansion of (\ref{eiko}) as a description
of successive interactions with the potential. Here, however, we have no 
potential, so that the identification of each term with successive pomeron
exchanges is not obvious.

The usual approach is to assume that the Taylor expansion of (\ref{eiko})
is such that the $n^{th}$ term corresponds to $n$-pomeron exchange.
In this case,
we take the eikonal form for the scattering amplitude.
\begin{eqnarray}
      A_e(s, t\!=\!0)
        \,=\, 2 \int \!d^2b\, [1-\exp(i\F_0(s,b))] \, .
\label{eik}   \nonumber
\end{eqnarray}

Before giving the results in this approach, a few comments are in order.
First of all, the eikonal is only a model. Indeed,
it is known \cite{LPcuts} that it does not reproduce properly the s-channel
cuts of the scattering amplitude coming
from multiple exchanges. For instance, already the second term, corresponding to
the two-pomeron cut, could have a suppression due to the structure of
the proton \cite{book}. 

Furthermore, the eikonal does not always guarantee unitarisation. Although
it obeys Eq. (\ref{unitarity}) for all values of $b$, it can produce,
after integration, amplitudes that violate the Froissart bound \cite{Froissart}, depending on the dependence of the form factor.
To see this, take the eikonal  in factorised form
\begin{eqnarray}
 \chi(s,b) = h(s) \ f(b),
\end{eqnarray}
with $h(s)=s^\Delta$,
and assume simple functional forms for the form factor $f(b)$, which allow an
analytical treatment.
If one considers a Gaussian form
\begin{eqnarray}
        f(b) \sim \exp(-b^2/R^2),
\end{eqnarray}
 one obtains
\begin{eqnarray}
  A(s,t=0) \sim i \ R^2 \ ( \Gamma(0,s^{\Delta}) \ + \ \gamma \ + \ \Delta \log{s} ),
\end{eqnarray}
 where
\begin{eqnarray}
  \Gamma(a,z) = \int_{z}^{\infty} \ t^{a-1} \  e^{-t} \ dt  \nonumber
\end{eqnarray}
 and, in our case,
\begin{eqnarray}
  \Gamma(0,s^{\Delta}) \rightarrow  \ 0, \ \ \
           s  \ \rightarrow \  \infty .
\end{eqnarray}
  If $R^2$ is independent from $s$, we have
\begin{eqnarray}
\sigma_{tot} \ \sim \  \log(s).
\end{eqnarray}
  whereas for $R^2$ growing like $ \log(s)$, we obtain
\begin{eqnarray}
\sigma_{tot} \  \sim  \ \log^2(s),
\end{eqnarray}
  so the total cross section respects the Froissart bound.
  However, if we take a polynomial form, such as that resulting from
dipole-dipole interactions  \cite{kozlov-dd}, 
\begin{eqnarray}
  f(b) \ \sim \frac{1}{b^4}
\end{eqnarray}
we obtain
\begin{eqnarray}
  A(s,t) \sim i \int_{0}^{\infty} \ \frac{ 1}{y \sqrt{y}}
  \ [1\ - \ \exp( - s^{\Delta} \ y)] \ = \ 2 \sqrt{\pi} \ s^{\Delta/2}.
\end{eqnarray}
with $y= 1/b^4$.  So, in this case, the scattering amplitude
does not satisfy the Froissart unitarity bound.
    This violation is not due to the divergence of $f(b)$ at small $b$. Indeed,
if we introduce an additional small constant radius $r$ which removes
the singular point $b=0$  in $f(b)$ and take
\begin{eqnarray}
  f(b) \ \sim \frac{1}{b^4+r^4}
\end{eqnarray}
 the answer, after some complicated algebra, is
\begin{eqnarray}
A(s,t=0)\ \sim \ \frac{1}{4r^2}[\pi s^{\Delta} \exp[-s^{\Delta}/(2r^4)]
  \ [I_{0}(s^{\Delta}/(2r^4)) +I_{1}(s^{\Delta}/ (2r^4))].
\end{eqnarray}
 The asymptotic value of the  Modified Bessel functions is
\begin{eqnarray}
  I_{0,1}(s^{\Delta}/(2r^4)) \ \sim \ \frac{r^2}{\sqrt{\pi} s^{\Delta/2}}.
\end{eqnarray}
Hence we again obtain for asymptotic high energies
\begin{eqnarray}
  A(s,t) \sim i \ \frac{\sqrt{\pi} \ s^{\Delta/2}}{2 \sqrt{2}}.
\end{eqnarray}
which again does not obey the Froissart bound.

So, in the following, we shall use the eikonal as a simple example. The form
factors that we use lead to a cross section that does respect the Froissart
bound. Furthermore, we shall be able to see whether the saturation effects
that we found depend on the picture of saturation, or depend only on
the onset of unitarising cuts.

We show in Fig.~\ref{profile} that, in this case, the profile function
never saturates: the cuts actually reduce the cross section from the start,
and never allow $\Gamma(s,b)$ to become larger than 1.
This means, as is shown in Fig.~\ref{sigtot} that eikonalisation will
give a suppression, even at lower energies. So one needs to modify
slightly the parameters of Table~1, to recover the low-energy fit. We find
that multiplying the couplings by 1.2 provides such an agreement,
and refer to this as the "renormalised eikonal", shown in Fig.~\ref{sigtot}.
We see that both curves are very close at the LHC, but exceed considerably
previous estimates \cite{compete}: whereas the total cross section at the
LHC was predicted to be $111.5\pm 1.2 \begin{array}{c} +4.1\\ -2.1 \end{array}$ mb, it is
now 152~mb.

In a similar manner, Fig.~\ref{elastic} shows that the elastic cross
section also gets large corrections from the cuts at lower energies. Again,
the renormalised eikonal is close to the saturated curve at low energy,
but this time deviates at higher energies, so that the ratio of
the elastic cross section to the total cross section changes by about
10\%.

\begin{figure}
\begin{center}
\mbox{\epsfxsize=70mm\epsffile{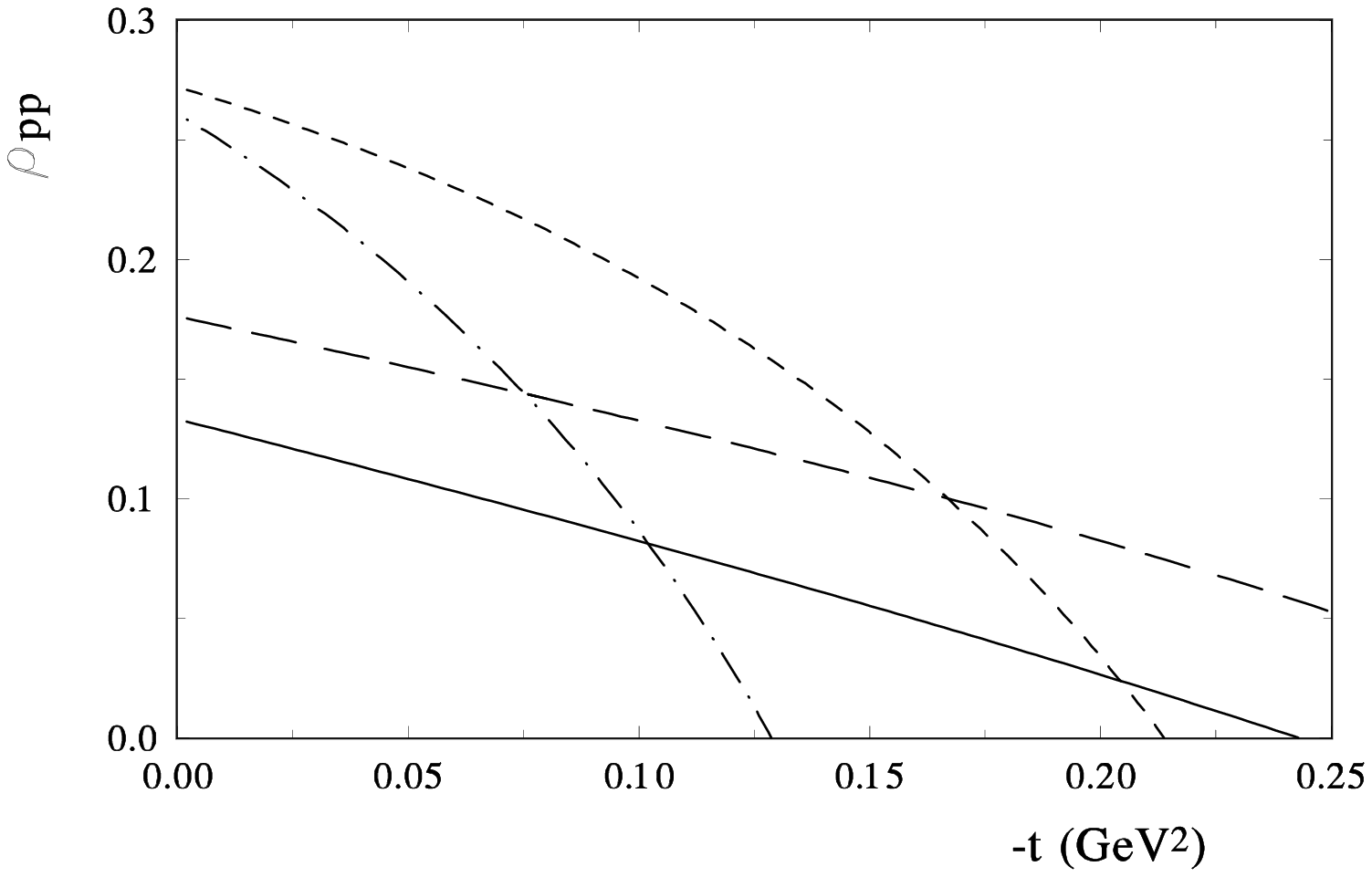}}
\mbox{\epsfxsize=70mm\epsffile{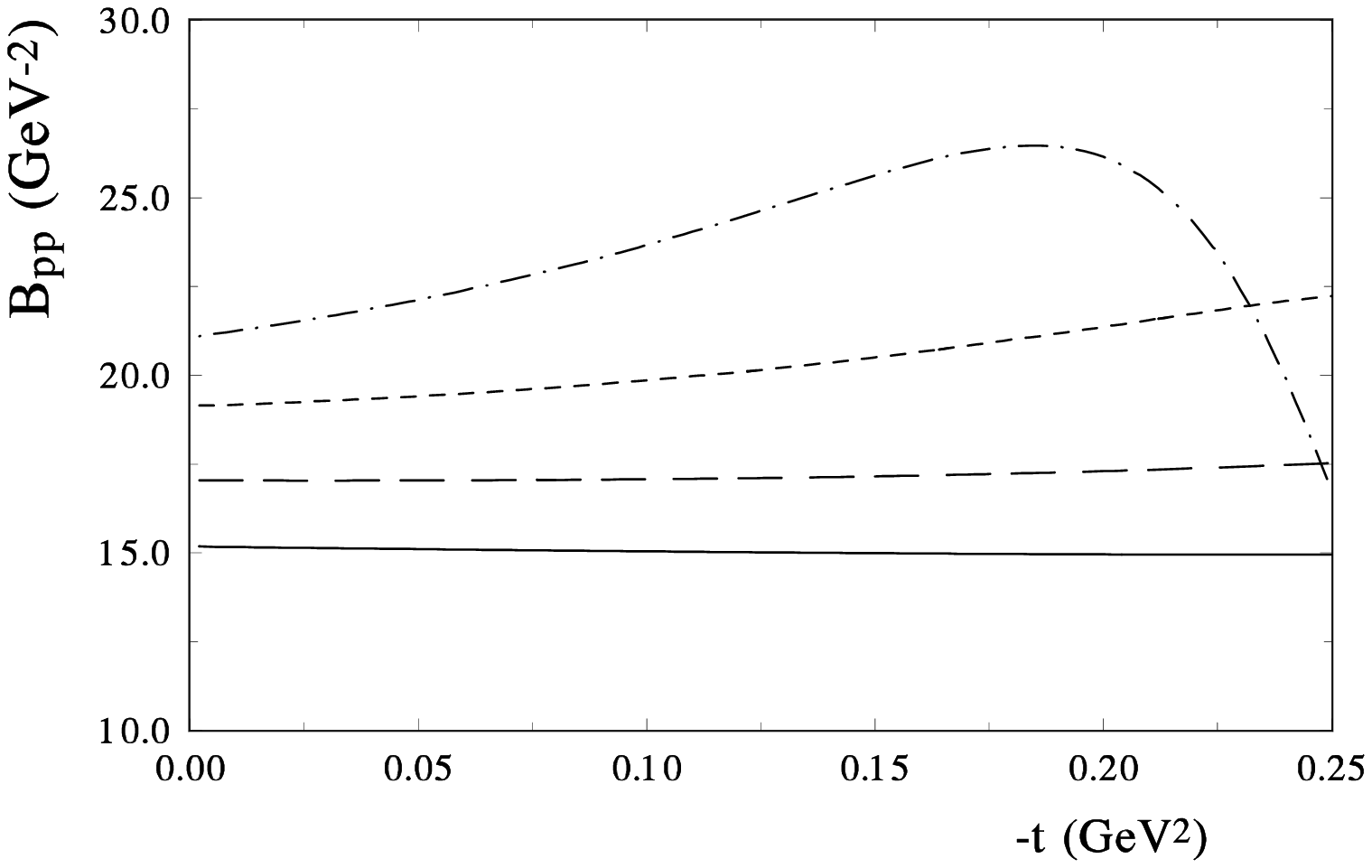}}
\end{center}
\caption{The ratio of the real to the imaginary part of the amplitude
(left), and the slope of the elastic differential cross section 
(right) as functions of $t$ 
for the eikonalised amplitude at various energies: 100 GeV (plain curve),
500 GeV (long dashes), 5 TeV (short dashes) and 14 TeV (dash-dotted curve).
}
\label{rhoeiko}
\end{figure}

We had seen that a striking feature of saturation was the behaviour
of the real part of the amplitude at small $t$, as the $\rho$ parameter
would have a drastic change of slope. We find that such an effect will also be
present in the eikonal case (see Fig.~\ref{rhoeiko}), although the real
part will be much bigger in this case. Similarly, we also show in 
Fig.~\ref{rhoeiko} the behaviour of the slope of the differential 
elastic cross section. Again, we see that it increases at small $t$, 
and then decreases towards the dip.

\section{Conclusion}
 Now the structure of the diffractive
 scattering amplitude cannot be obtained from  first principles
 or from QCD. The procedure used to extract
 such structure and parameters of the elastic scattering amplitude
 from the experimental data requires some different assumptions.
If we assume that the hard pomeron is present in soft data, it will
lead to a cross section of the order of 
150 mb (a similar conclusion, via a different argument, has been 
reached in \cite{peter}). The uncertainty in this number 
is quite large, as unitarisation and saturation schemes are numerous. 
Hence it seems that the total cross section can be anywhere between
108 mb \cite{compete} and 150 mb. 

Other observables may be used to decide whether one has a simple
extrapolation of the lower-energy data, such as in \cite{compete},
or one is entering a new regime of unitarisation. Indeed,
in the presence of the hard Pomeron,
the saturation effects, which must then be present at LHC energies, 
  can change the behavior of the real part of the cross section making
it smaller than expected, especially in the near-forward region,
and of the slope of the differential elastic scattering cross section.

Despite the lack of an absolute prediction for total cross sections,
the observation of such features would be a clear sign that a new regime
of strong interactions has been reached.
\section*{Acknowledgements}
O.V.S. acknowledges the support of FNRS (Belgium) for visits to the University
of Li\`ege where part of this work was done.
We thank E. Martynov, S. Lengyel, G. Soyez and P.V. Landshoff for discussions.

\end{document}